\documentclass{aip-cp}

\usepackage[numbers]{natbib}
\usepackage{rotating}
\usepackage{graphicx}
\usepackage{multirow}
\usepackage{subfigure}

\newcommand{\maddm}{MadDM}
\newcommand{\madgraph}{MadGraph5$\underline{\hspace{0.16cm} }$aMC@NLO{}}

\newcommand {\beq} {\begin{equation}}
\newcommand {\eeq} {\end{equation}}
\newcommand {\bea} {\begin{eqnarray}}
\newcommand {\eea} {\end{eqnarray}}

\begin{document}

\title{MadDM: New Dark Matter Tool in the LHC era}

\author[aff1]{Mihailo Backovi\'{c}}
\author[aff1]{Antony Martini}
\author[aff2]{Kyoungchul Kong\corref{cor1}}
\author[aff3]{Olivier Mattelaer}
\author[aff2]{Gopolang Mohlabeng}

\affil[aff1]{Centre for Cosmology, Particle Physics and Phenomenology (CP3), \\ Universit\'{e} catholique de Louvain,  Chemin du Cyclotron 2, B-1348 Louvain-la-Neuve, Belgium}
\affil[aff2]{Department of Physics and Astronomy, University of Kansas, Lawrence, KS, 66045, USA}
\affil[aff3]{Institute for Particle Physics Phenomenology (IPPP), Durham University, Durham, DH1 3 LF, United Kingdom}
\corresp[cor1]{A talk given by K. Kong at PPC2015 conference, Deadwood, SD}

\maketitle

\begin{abstract}
We present the updated version of MadDM, a new dark matter tool based on \madgraph{} framework.
New version includes direct detection capability in addition to relic abundance computation.
In this article, we provide short description of the implementation of relevant effective operators and validations against existing results in literature.
\end{abstract}

 \section{Introduction}

Diversity of dark matter candidates requires the balanced experimental program in direct/indirect detection, astroparticle probes and collider searches. Often these approaches make heavy use of numerical tools and computer simulations, and especially 
there is a variety of useful programs for the purpose of collider searches.
In the case of dark matter, there are a few tools available but practically micrOMEGAs \cite{Belanger:2008sj} is the only tool that works for any arbitrary model for physics beyond the Standard Model. It is based on CalcHEP and includes various components such as relic abundance, direct and indirect detection capabilities. 

In this talk we introduce the updated version of MadDM \cite{Backovic:2015cra}, a new dark matter tool based on a well known collider package, \madgraph{} framework \cite{Alwall:2011uj}.
New version incorporates direct detection capability in addition to relic abundance calculation \cite{Backovic:2013dpa}.
Short description on the implementation of relevant effective operators and validations against existing results are presented in the following sections.

\section{MadDM and Validation}

MadDM is available for download from Launchpad web site , https://launchpad.net/maddm and no separate installation is required.
However it uses python routines in \madgraph{} and MadDM needs to be copied under the main directory of \madgraph{}.
New version of MadDM inherits the already established structure of MadDM v1. 
A Python module prepares relevant amplitudes for relic abundance and direct detection (with ALOHA \cite{deAquino:2011ub}), 
while a FORTRAN module deals with the numerical calculations. 
MadDM is compatible with any UFO (Universal FeynRules Output \cite{Degrande:2011ua}) model that contains a dark matter candidate and can be easily linked to any width or mass spectrum generator which can produce a parameter card in Les Houches format.

For direct detection, MadDM follows the procedure described in Ref. \cite{Belanger:2008sj}. 
At the quark level, quark-DM interaction at the low energy is written in terms of a set of effective operators in Table \ref{EffTab}.
They are further classified as odd and even operators under the interchanges of quark and anti-quarks. 
Projection over the effective operators select either spin-independent (SI) and spin-dependent (SD) contributions.
Finally the corresponding cross sections are computed with the user's choice of target material and several options such as 
exposure time, size of the target material, threshold cut, detector resolution etc.
\begin{table}[tb]
  \centering
\scalebox{0.9}{
  \begin{tabular}{|c|c|c|c|}
  \hline
   & DM spin & Even & Odd \\
  \hline
  \multirow{4}{*}{SI} & & scalar current & vector current \\
  & $0$ & $2M_{\chi} S S^{*} \bar{\psi}_q \psi_q $ & $i \left( \partial_{\mu} S \, S^{*} - S \partial_{\mu} S^{*} \right) \bar{\psi}_q \gamma^{\mu} \psi_q$ \\
  & $1/2$ & $\bar{\psi}_{\chi} \psi_{\chi} \bar{\psi}_q \psi_q$ & $\bar{\psi}_{\chi} \gamma_{\mu} \psi_{\chi} \bar{\psi}_q \gamma^{\mu} \psi_q$ \\
  & $1$ & $2M_{\chi} A_{\chi \mu}^{*} A_{\chi}^{\mu} \bar{\psi}_q \psi_q$ & $i \left( A_{\chi}^{* \alpha} \partial_{\mu} A_{\chi \alpha} - A_{\chi}^{\alpha} \partial_{\mu} A_{\chi \alpha}^{*} \right) \bar{\psi}_q \gamma_{\mu} \psi_q $  \\
  \hline
  \hline
  \multirow{4}{*}{SD} & & axial-vector current & tensor current \\
  & $1/2$ & $\bar{\psi}_{\chi} \gamma^{\mu} \gamma^5 \psi_{\chi} \bar{\psi}_{q} \gamma_{\mu} \gamma_5 \psi_{q}$ & $- \frac12 \bar{\psi}_{\chi} \sigma_{\mu \nu} \psi_{\chi} \bar{\psi}_{q} \sigma^{\mu \nu} \psi_{q}$ \\
  & $1$ & $ \quad \sqrt{6} \left( \partial_{\alpha} A_{\chi \beta}^{*} A_{\chi \nu} - A_{\chi \beta}^* \partial_{\alpha} A_{\chi \nu} \right) \epsilon^{\alpha \beta \nu \mu} \bar{\psi}_q \gamma_5 \gamma_{\mu} \psi_q \quad $ & $ \quad i \frac{\sqrt{3}}{2} \left( A_{\chi \mu} A_{\chi \nu}^* - A_{\chi \mu}^* A_{\chi \nu} \right) \bar{\psi}_q \sigma^{\mu \nu} \bar{\psi}_q \quad $  \\
  \hline
  \end{tabular}
  \caption{List of effective operators taken from Ref. \cite{Belanger:2008sj}, implemented into \maddm{}. $S$, $\psi_{\chi}$ and $A_{\chi}$ correspond to scalar, fermion and vector DM fields, respectively.  }
  \label{EffTab} }
\end{table}

\begin{figure}[t]
\centerline{
\includegraphics[scale=0.33]{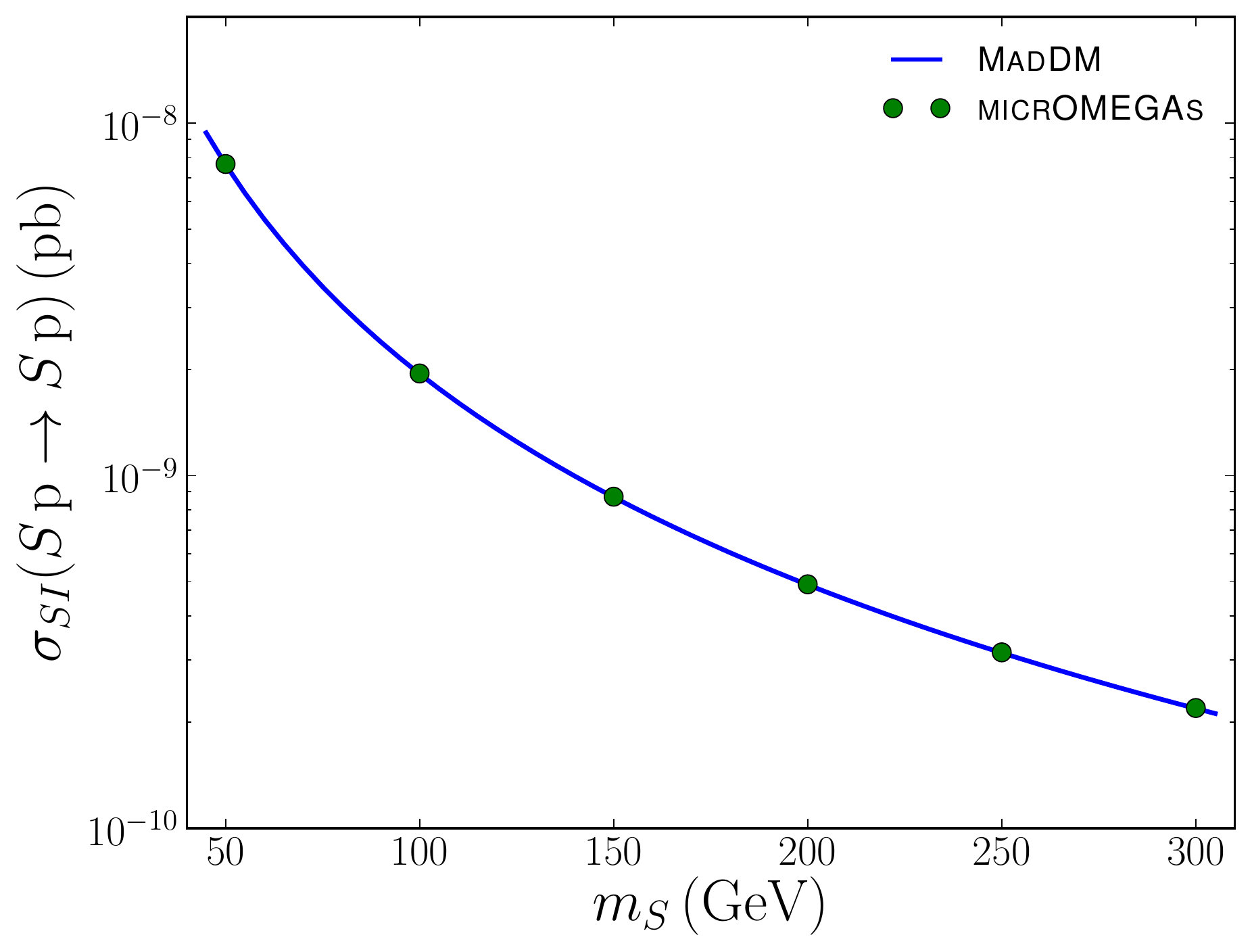}\hspace{1cm}
 \includegraphics[scale=0.33]{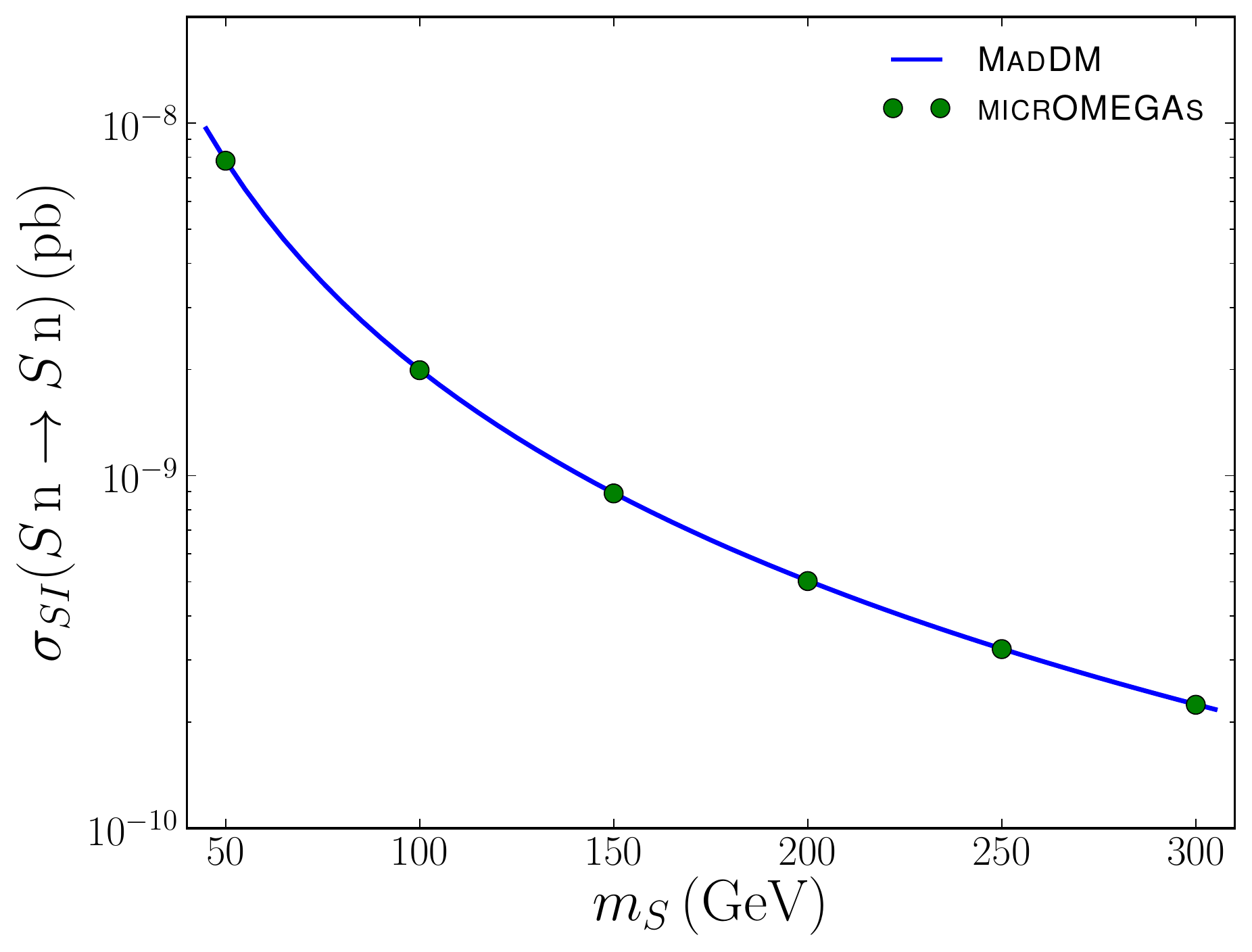}    }    %
 \caption{Spin-independent elastic scattering cross section of scalar DM in the simplified SM model scenario with proton and neutron for $\delta = 0.1$.}
 \label{SimpModels1}
\end{figure}

\begin{figure}[h]
\centerline{
 \includegraphics[scale=0.33]{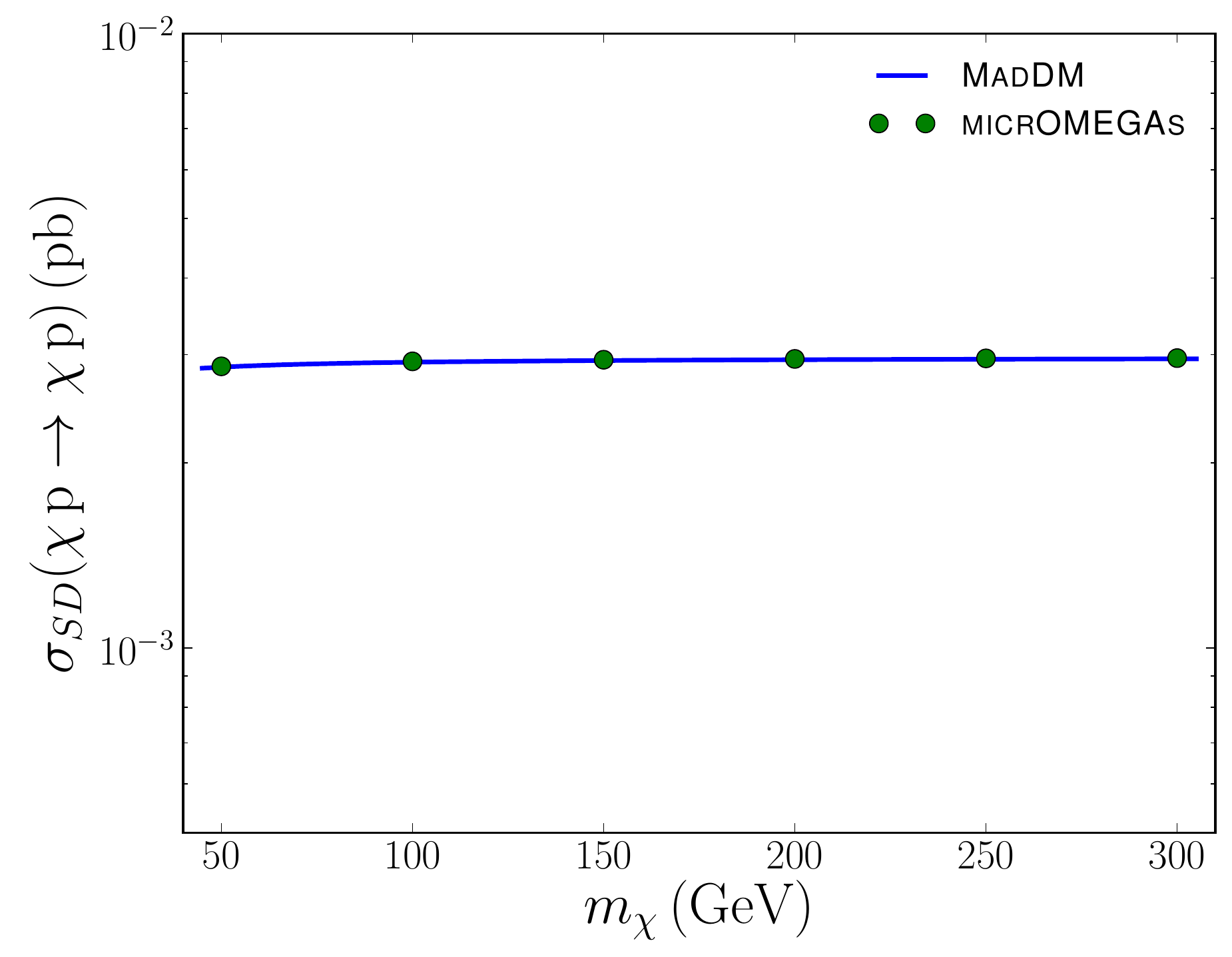}   \hspace{1cm}
 \includegraphics[scale=0.33]{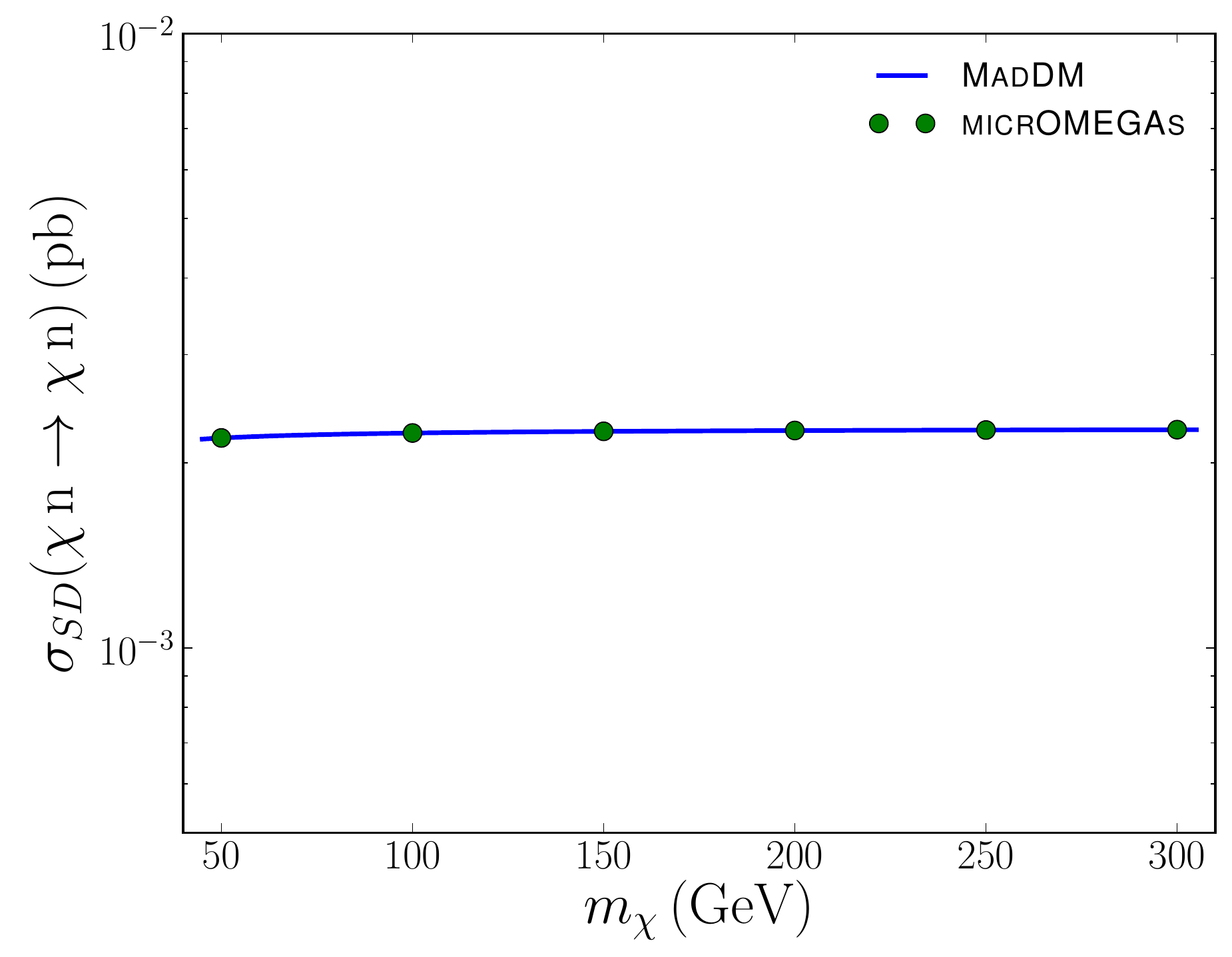}    } 
 \caption{Spin-dependent elastic scattering cross section of Dirac fermion DM with proton and neutron for $g_{\chi}^A = 0.1$.}
 \label{SimpModels2}
\end{figure}

We have made various consistency checks against existing results in literature.
First we have considered all possible simplified models with different mediators (scalar, fermion and vector) and different DM candidate (scalar, fermion and vector), which are shown in Table \ref{EffTab} and verified that SI and SD cross sections from \maddm{} are consistent with results from micrOMEGAs. 
As an illustration of validation, here we consider two examples: scalar DM ($S$) with Higgs ($H$) as a mediator ($\frac{\delta}{2} H^\dagger H S$) and Dirac fermion DM ($\chi$) with a vector ($V_\mu$) mediator ($g_{\chi}^A \, \bar{\chi} \gamma^{\mu} \gamma_5 \chi V_{\mu} + g_q^A \, \bar{q} \gamma^{\mu} \gamma_5 q V_{\mu}$).
SI and SD elastic scattering cross sections are shown in Fig. \ref{SimpModels1} and Fig. \ref{SimpModels2}, respectively, together with 
results obtained using micrOMEGAs. In both cases, we find an excellent agreement between \maddm{} and micrOMEGAs with the differences at $\lesssim 1 \%$ level.
Similar results are obtained with all operators in the Table \ref{EffTab}.

\begin{figure}[t]
\centering
 \includegraphics[scale=0.33]{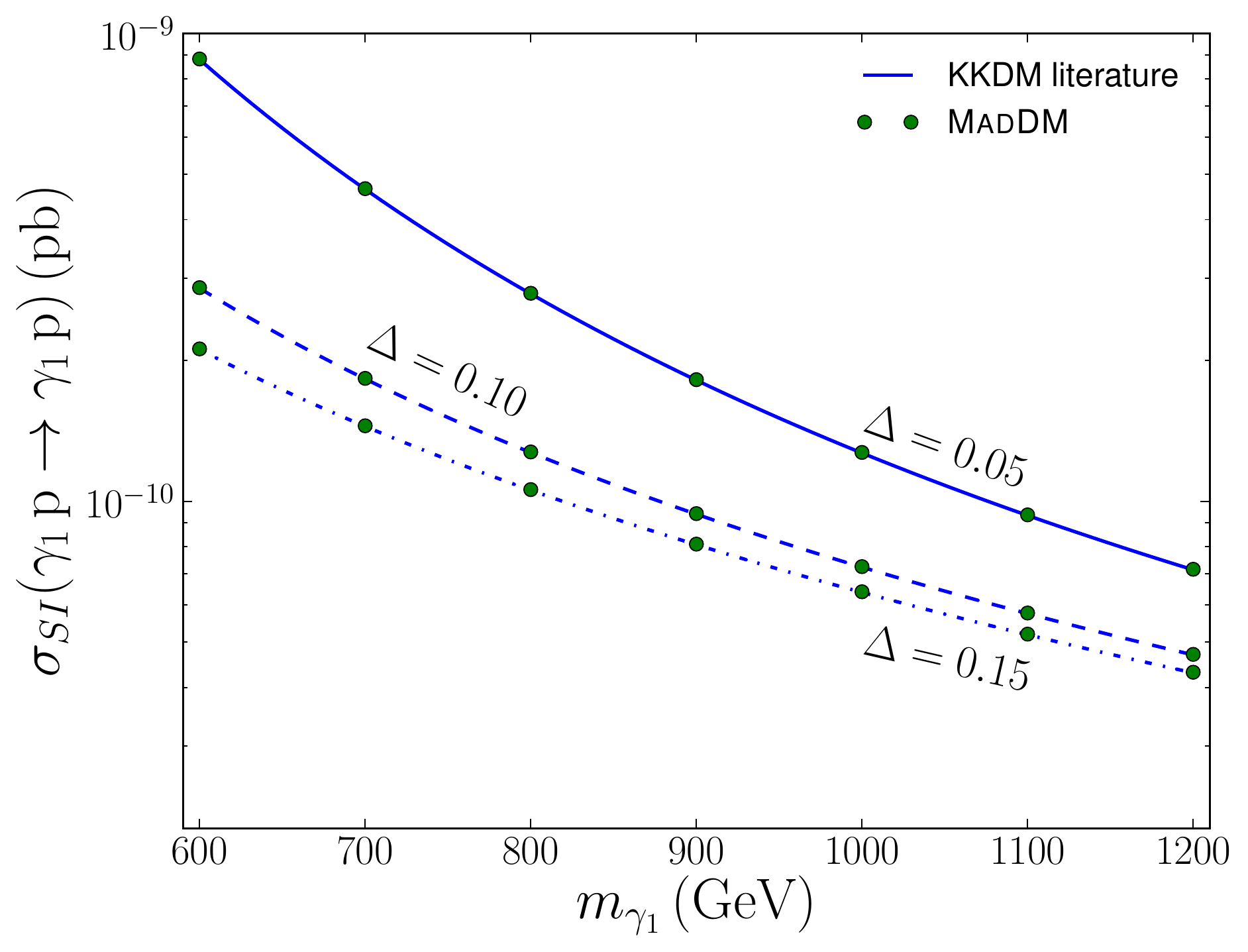}   \hspace{1cm}
 \includegraphics[scale=0.33]{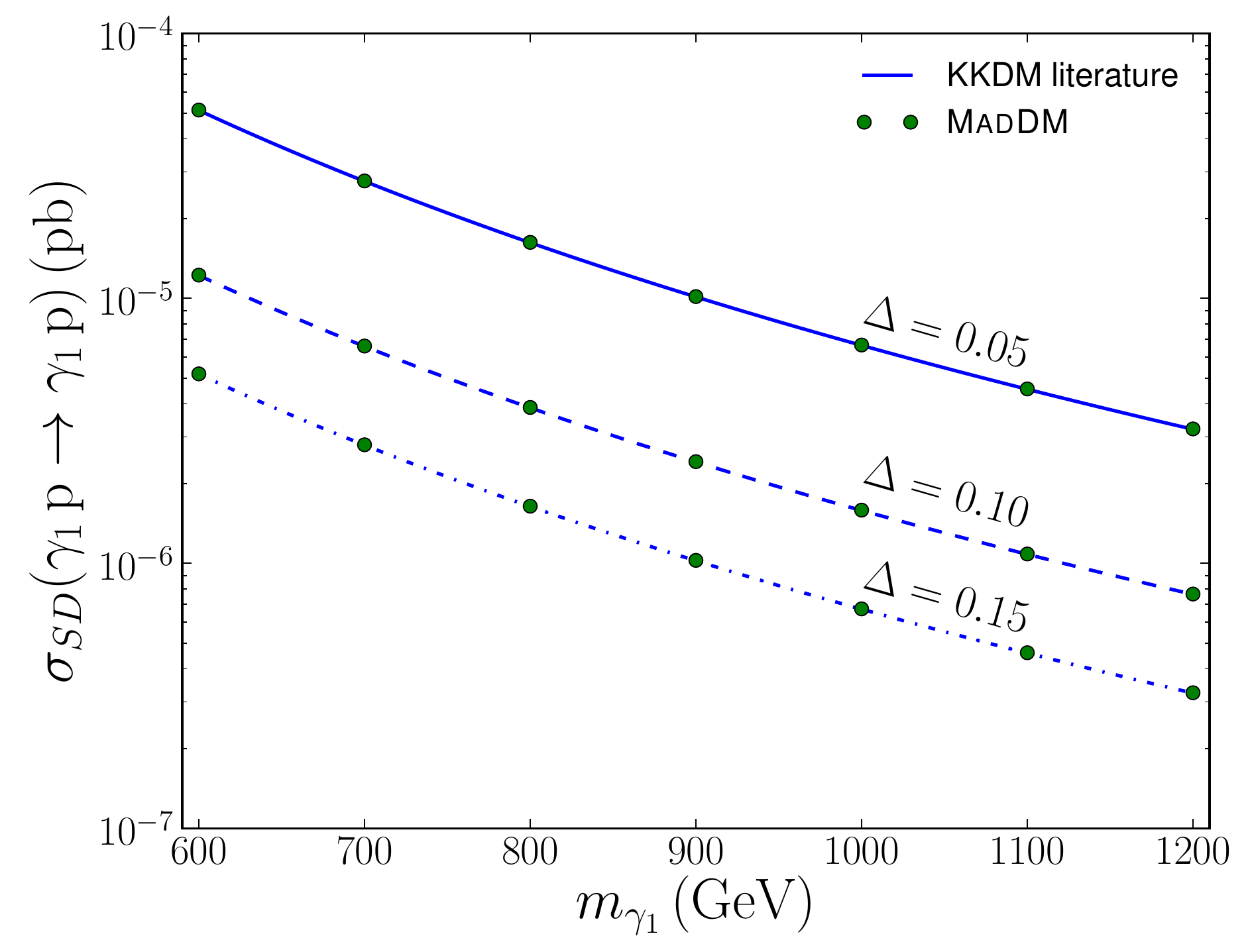} 
 \caption{Spin-independent (left panel) and spin-dependent (right panel) DM-proton cross section for MUED. The blue curves correspond to theoretical values coming from Kaluza-Klein dark matter literature \cite{Arrenberg:2008wy,Cheng:2002ej,Servant:2002hb}. The green points are the MadDM data. We show the results for three values of the mass splitting parameter, $\Delta$.}
 \label{ued}
\end{figure}

As for more complete dark matter models, we have considered Minimal Universal Extra Dimensions (MUED) and Minimal Supersymmetric Standard Model (MSSM) with SPS1a mass spectrum. For MSSM, 
Two cross sections from micrOMEGAs and MadDM agree at the $\sim 1$\% level. We have checked that this is true if we vary the neutralino mass around the SPS1a point as well. For MUED, Fig. \ref{ued} shows excellent agreement between MadDM cross sections and results in literature \cite{Arrenberg:2008wy,Cheng:2002ej,Servant:2002hb}.

%
%

\begin{figure}[t]
\centerline{
\hspace{0.6cm}
\includegraphics[scale=0.32]{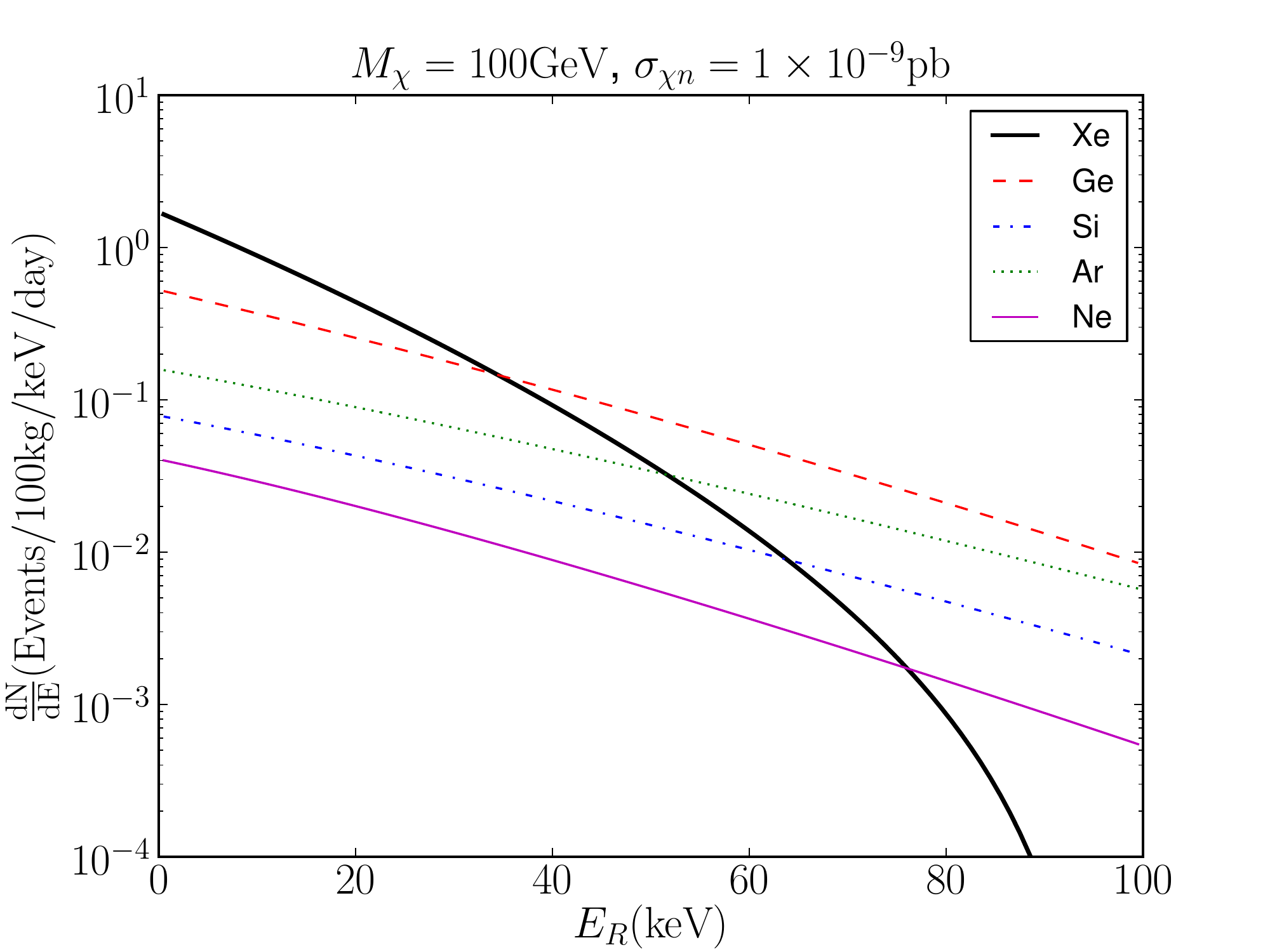}
\hspace{0.8cm}
\includegraphics[scale=0.32]{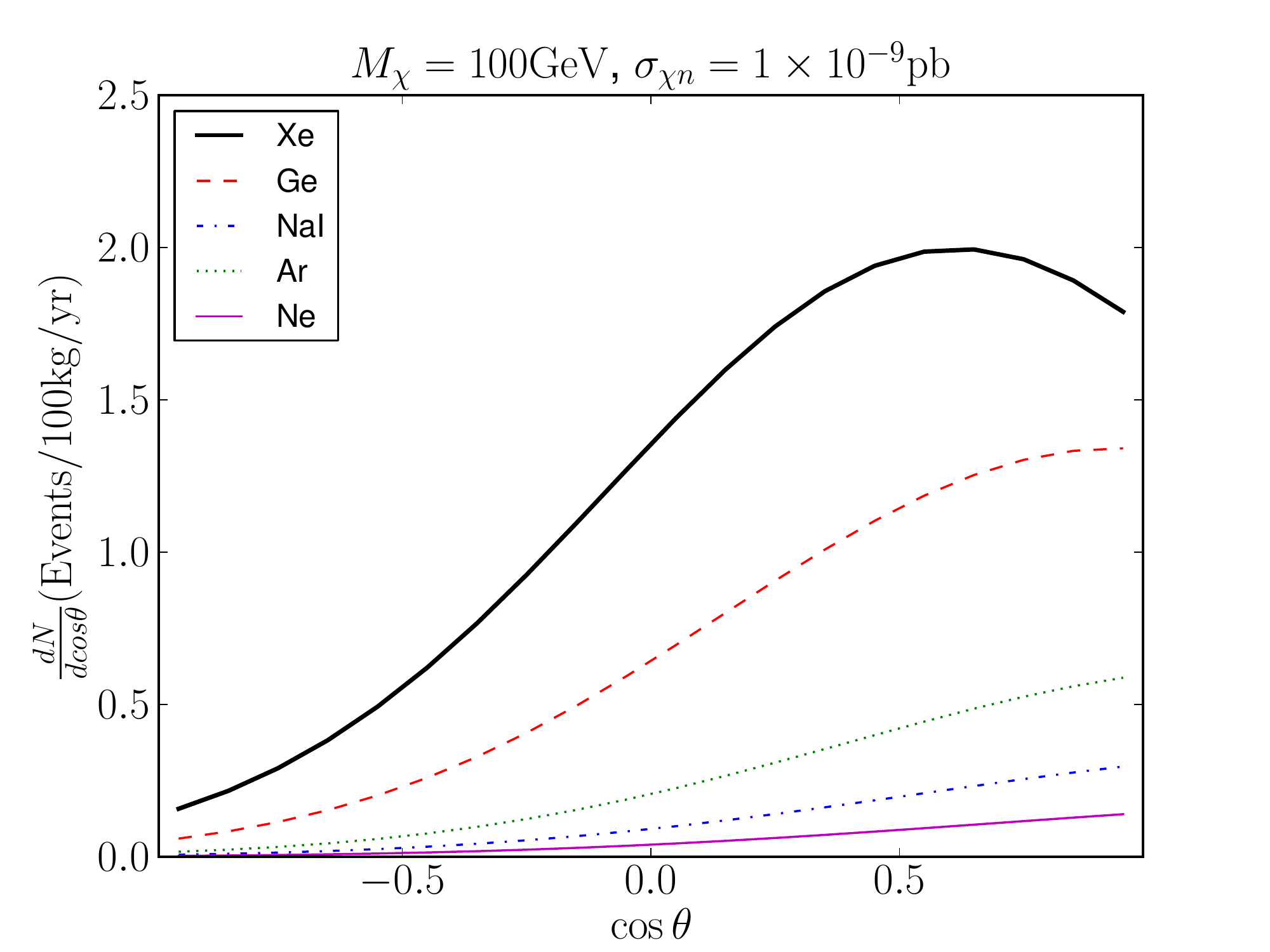} }
\caption{Nuclear recoil energy (left panel) and angular (right panel) distributions for spin-independent interactions for different materials, assuming a 100 kg detector measuring events over one year for a DM mass of 100 GeV and DM-nucleon cross-section of $1\times 10^{-9}$ pb.  \label{dndetarek}}
\end{figure}

Upon validating the DM-nucleon scattering cross section in \maddm{},  we calculated the recoil rates for DM scattering off a target nucleus.  
We performed a simple, model independent validation of the recoil rate calculation, where we simply assumed that the DM-nucleon cross section $\sigma_{\chi n} = 10^9$ pb, chosen for the purpose of comparison with the results from Ref. \cite{Saab:2014lda}.
To reproduce the SI recoil rates as a function of energy/angle as in Ref. \cite{Saab:2014lda}, we employed the differential recoil spectrum  integrated over time and angle/energy. Fig. \ref{dndetarek} shows the spin-independent recoil rates as a function of recoil energy (left) and recoil angle (right). We find that both distributions are in a very good agreement with the results found in Ref. \cite{Saab:2014lda}, over a wide range of target materials.
%

\begin{figure}[t]
\centerline{
\includegraphics[scale=0.38]{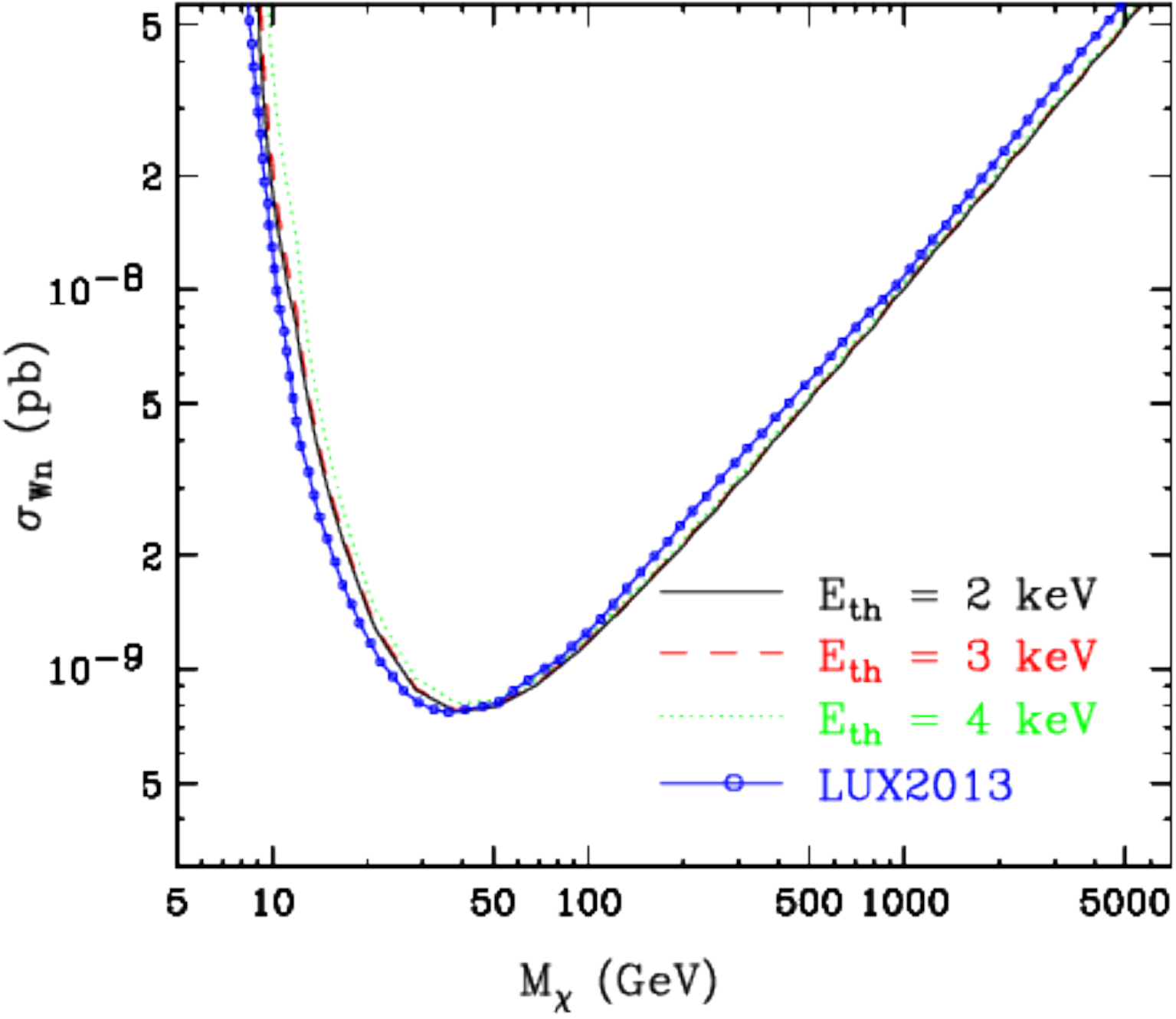}
\hspace{1cm}
\includegraphics[scale=0.38]{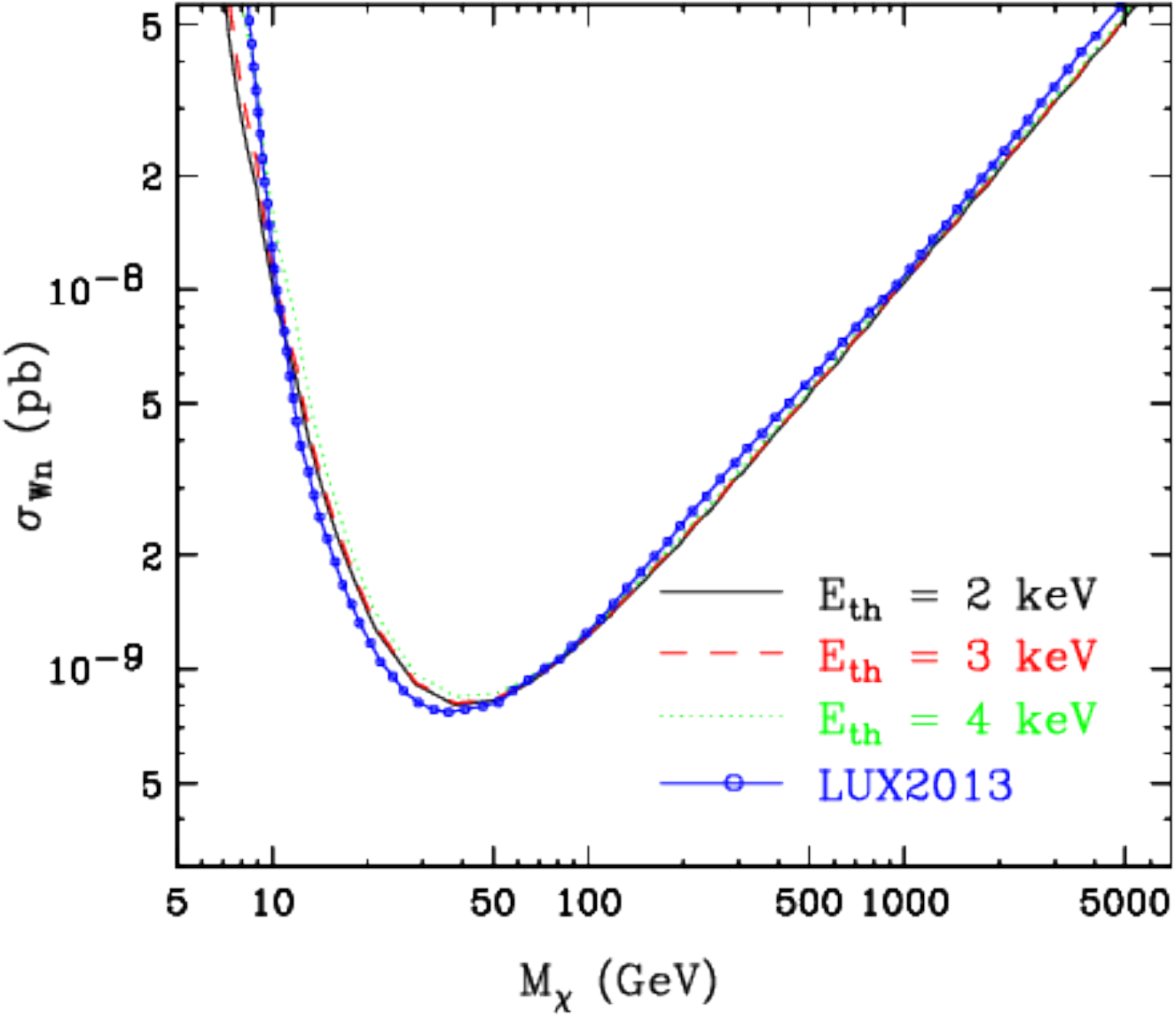} }
\caption{ 90\% confidence limits on the spin-independent DM-nucleon scattering cross-section (in picobarns) 
for an unsmeared energy distribution  (left panel) and the smeared distribution with $\lambda = 1$ (right panel). 
Limits are obtained from \maddm{} for 2 keV (black solid), 3 keV (red dashed) and 4 keV (green dot-dashed) and 
LUX limits are shown in blue curve with circular data points.}
\label{lux_rep}
\end{figure}

 
As a final validation of the \maddm{}, we have reproduced the exclusion limit on the DM-nucleon scattering cross-section as a function of DM mass as in the LUX 2013 paper \cite{Akerib:2013tjd}. We assumed the efficiency function of nuclear recoils displayed in the black curve of Fig. 1 in Ref. \cite{Akerib:2013tjd}.  
Exclusion curves shown in Fig. \ref{lux_rep} are obtained assuming 2.3 events, which coincide with the number of events at 90\% confidence as required by the Feldman-Cousins confidence intervals. We find a good agreement between the LUX data and limits from \maddm{}. 
The exclusion curves can be obtained in \maddm{} by using \verb|LUX_Exclusion| routine found in the test routines part in \verb|maddm.f|. The routine multiplies $\frac{dR}{dE}$ by the efficiency obtained from Ref. \cite{Akerib:2013tjd}, which is is then weighted by a 50\% acceptance rate for nuclear recoils as stated in the LUX analysis.  From the recoil spectrum weighted by the efficiency and the acceptance rate, the function then calculates the total number of expected events. 
The default value for the detector efficiency is 100\%, and can be easily replaced with a user defined function.

\section{Outlook}

Future version of MadDM would include various features such as 
improvement on precision and speed in computation of relic abundance, 
relic abundance beyond the leading order, 
direct detection with loop-induced operators,
indirect detection, web version, etc. 
Furthermore various dark matter models with more complicated dark sector such as one in Ref. \cite{Kong:2014mia} will be treated properly.

\vspace{-0.7cm}
\section{ACKNOWLEDGMENTS}
We thank the Center for Theoretical Underground Physics and Related Areas (CETUP* 2015) for hospitality and partial support during Dark Matter workshop and PPC2015 conference.
Especially we are grateful to Barbara Szczerbinska for all the arrangements and encouragement.
This work is supported in part by the Belgian Federal Science Policy Office through the Interuniversity Attraction Pole P7/37, by the National Research Foundation of South Africa under Grant No. 88614, and by the U.S. DOE under Grant No. DE-FG02-12ER41809, and by Durham International Junior Research Fellowships.



\end{document}